# Multi-layered chalcogenides with potential for magnetism and superconductivity

Li Li,[1] David S. Parker,[1] Clarina R. dela Cruz,[2] and Athena S. Sefat[1]

[1]*Materials Science & Technology Division, Oak Ridge National Laboratory, Oak Ridge, TN 37831*
[2] *Quantum Condensed Matter Division, Oak Ridge National Laboratory, Oak Ridge, TN 37831*

**Abstract**

Layered thallium copper chalcogenides can form single, double, or triple layers of Cu-*Ch* separated by Tl sheets. Here we report on the preparation and properties of Tl-based materials of $TlCu_2Se_2$, $TlCu_4S_3$, $TlCu_4Se_3$ and $TlCu_6S_4$. Having no long-range magnetism for these materials is quite surprising considering the possibilities of inter- and intra-layer exchange interactions through Cu 3*d*, and we measure by magnetic susceptibility and confirm by neutron diffraction. First principles density-functional theory calculations for both the single-layer $TlCu_2Se_2$ (isostructural to the '122' iron-based superconductors) and the double-layer $TlCu_4Se_3$ suggest a lack of Fermi-level spectral weight that is needed to drive a magnetic or superconducting instability. However, for multiple structural layers with Fe, there is much greater likelihood for magnetism and superconductivity.

**Introduction**

Owing to 2D crystal structures and electronic states, layered materials are known to exhibit many anomalous, competing, coexisting, or intertwined electronic, magnetic properties, and structural order parameters, which have promising functional materials applications. One of the most mysterious and elusive properties in condensed-matter physics remains as unconventional electron pairing in high-temperature superconductors (HTS). The two famous families of HTS of cuprates and Fe-based superconductors (FeSC) share structural layers, with square-planar Cu or Fe sheets, with low-dimensional spin and charge correlations [1, 2, 3, 4]. Stimulated by the discovery of HTS, tremendous efforts have been made to synthesize similar layered materials, and to understand crystal structure, chemical inhomogeneity, disorder, inter-site exchanges, and composition relations [5-11]. Among the layered superconductors, the chalcogenides are one of the notable groups, because they generally possess a van der Waals-like stacking structure for which ions can be easily intercalated into the interlayer sites, changing the physical properties. For the non-oxide transition-metal dichalcogenide compounds of M*Ch*$_2$ (M = Ti, Ta, Nb or Ir), superconductivity shows the competition, coexistence and even cooperation with charge-density-wave (CDW) states or phase transition [12-15]. By the intercalation of Cu or Sr in the $Bi_2Se_3$ topological insulator, bulk superconductivity can be achieved in single crystals [16-17]. Superconductivity was discovered in the $BiS_2$-based layered compounds by electron doping into the conduction layers [18]. Moreover, superconductivity is found in the simple binaries of FeSe and FeS [19], even in those complex compounds containing intercalations between FeSe layers, such as *Am*$Fe_2Se_2$ (*Am* = alkali metals) and $Li_x(NH_2)_y(NH_3)_{1-y}Fe_2Se_2$ [20]. Many of the copper chalcogenides show various unique properties and exhibit a large variety of crystal structures and phase transitions [21, 22]. Low-$T_c$ superconductivity have been observed in copper dichalcogenides Cu*Ch*$_2$, and spinels of $CuRh_2S_4$ and $CuRh_2Se_4$ [23, 24].

Although HTS in cuprates and FeSC are only explored in crystal structures with 'single' layers of Cu or Fe, there are crystal structures that show multiple transition metal layers, which are harder to stabilize in bulk forms. For example, there is $TlCu_{2n}Ch_{n+1}$ series, with Cu tetrahedral sharing edges similar to iron arsenides. The thickness of the $Cu_{2n}Ch_{n+1}$ layers increases as *n* increases, and the structural variations of single, double, and triple layers are shown in **Fig. 1**. Cu*Ch*$_4$ units share edges not only in the plane of the layers but also along the direction perpendicular to the plane of the layers along *c*-axis. An alternative way to describe the Cu-$Ch$ slab is to consider a square net of Cu atoms sandwiched in a staggered way by square nets of $Ch$ atoms; Tl cations are surrounded by a cube of $Ch$ atoms. TlCu$_2$Ch$_2$ (n=1) adopts the ThCr$_2$Si$_2$ structure (space group $I4/mmm$) [25, 26]. In this crystal structure, Cu-$Ch$ layers are isolated from each other by Tl sheets. For TlCu$_4$Ch$_3$ (n=2) CuCh$_4$ now form thicker triple Cu$_4$Ch$_3$ layers sandwiched by the metal sheets [27,28]. TlCu$_6$Ch$_4$ (n=3) shows thickest copper-chalcogenide slabs of Cu$_6$Ch$_4$ layers [29]. For n=2 and n=3, one would suspect a variety of inter- and intra-layer exchange interactions through any transition metals with unpaired 3$d$ electrons. Although this manuscript focuses on Cu, one envisions similar structures with Fe, Co, Ni, etc, although they may be more difficult to stabilize in bulk forms.

This manuscript reports on properties of layered Tl-Cu-$Ch$ and compare to reports of $A$Cu$_{2n}$Ch$_{n+1}$ with $A$ = Ba, K, Rb, Cs, and $Ch$ = S, Se. Because it was not possible to synthesize these layered structures with Fe instead of Cu in bulk forms, we report electronic structure calculations for the single- and double-layered Tl-Fe-Se structures. Although all Cu-based materials are found to show metallic behavior, lacking magnetism and superconductivity, the Fe analogs should have much higher likelihood for magnetism.

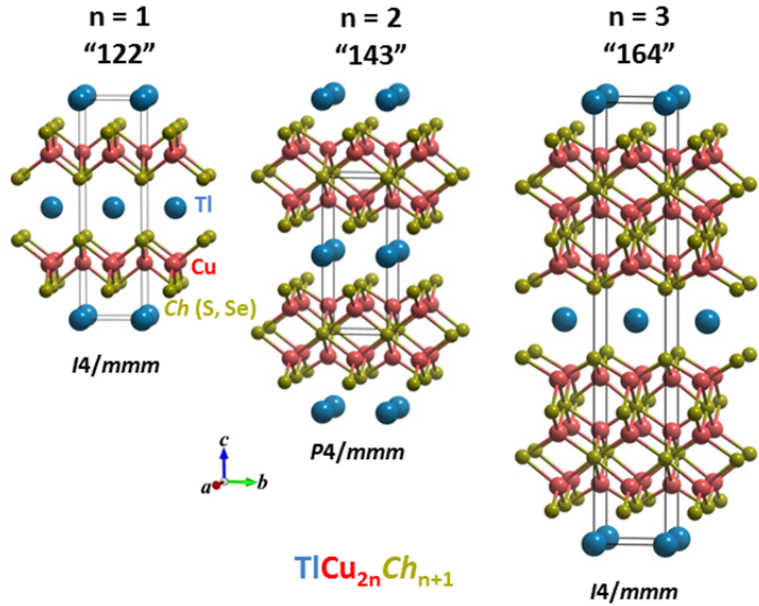

**Fig. 1.** Layered structural variation of copper-chalcogenide ternaries, from n=1 and n=2, to triple layers (n=3).

**Experimental**

There are previous synthesis reports on "122" [25, 26], "143" [27, 28] and "164" [29]; however, we were unable to replicate them for pure products and here we describe our procedure. TlCu$_2$Se$_2$, TlCu$_4$S$_3$, TlCu$_6$S$_4$ and TlCu$_4$Se$_3$ were synthesized from Tl chunks (99.999%), Cu powders (99.999%) and S or Se shot (99.999%), which were stored and handled in a helium-filled glovebox with the moisture and oxygen below 0.1 ppm. Prior to its use, copper powder was reduced in a diluted H$_2$ stream (5% H$_2$/Ar) for 12 h. For the triple-layered TlCu$_6$S$_4$, the starting materials were combined in stoichiometric ratio (~ 3 g total mass), put in 5cc alumina crucibles, then sealed inside silica ampoules evacuated to 10$^{-2}$ Torr. The ampoule was slowly heated up to 350 °C in a box furnace, kept for 12 h, and then furnace-cooled to room temperature. The products were ground and pelletized in air, sealed inside evacuated silica ampoules, and annealed at 350 °C for another 12 h. After furnace cooling, one more circle of grinding, pelletizing, and annealing (12 h at 350 °C) was required for pure phase. It should be addressed that the phase purity is very hard to obtain. We spent months to reduce the amount of impurity and improve our growth condition by adjusting starting materials ratio, ampoules size, degree of vacuum, and annealing temperature. Just 25 °C difference will generate impure phases, e.g. annealing at 325 °C will generate mixtures of $TlCu_4S_3$ (52.5%), $TlCu_2S_2$ (30.5%) and $CuS$ (17.0%), or mixtures of $TlCu_4S_3$ (90.5%), $CuS$ (5.3%), and $Cu_9S_5$ (4.2%) will come out after annealing at 375 °C. We also tried to synthesize $TlCu_6Se_4$, $TlCu_6Te_4$, $TlFe_6S_4$, $TlCo_6S_4$, and $TlNi_6S_4$, no phase pure products were obtained. Benefiting from the synthesis procedure of $TlCu_6S_4$ just explained, the synthesis of single-layered $TlCu_2Se_2$, and doubled-layered $TlCu_4S_3$ and $TlCu_4Se_3$ structures went smoothly. In fact, similar procedure was utilized, except that increasing annealing temperature to 400 °C was required. All the products were black and stable in air.

Powder X-ray diffraction (XRD) patterns were collected at room temperature by a PANalytical X'Pert PRO MPD X-ray powder diffractometer with Cu Kα radiation. There are no impurity phases detected for $TlCu_6S_4$ and $TlCu_4Se_3$. Only trace amount of impurity phases are present in the XRD patterns of $TlCu_2Se_2$ and $TlCu_4S_3$ (less than 2% by mass); impurity phase was estimated using the Hill and Howard method [30]. For $TlCu_2Se_2$, the only impurity detected was $TlCu_3Se_2$, while for $TlCu_2Se_2$, it was $TlCuS_2$. Temperature-dependent electrical and thermal transport measurements were performed in a Quantum Design (QD) Physical Property Measurement System (PPMS), using a standard four-probe method. Temperature-dependent magnetization measurements were performed in a QD Magnetic Property Measurement System (MPMS). The neutron diffraction measurements were done at the HB2a High Resolution Powder Diffractometer housed at the High Flux Isotope Reactor, ORNL. The patterns were measured using a wavelength of 2.411 Å with pre-monochromator, pre-sample and pre-detector collimation of open-21'-12'. A top loading low temperature cryostat was used to measure the diffraction pattern between 250 K and 4 K. First principles density-functional theory calculations of the electronic structure of $TlT_2Se_2$ and $TlT_4Se_3$ ($T$ = Cu or Fe) were performed using the generalized gradient approximation (GGA [31]) as implemented in the plane-wave all-electron code WIEN2K [32].

**Results and Discussions**

**Fig. 2**, top, shows the temperature dependence of electrical resistivity, ρ(T), on polycrystalline $TlCu_2Se_2$, $TlCu_4S_3$, $TlCu_4Se_3$ and $TlCu_6S_4$ samples. All of them exhibit metallic behavior with ρ increasing with temperature. They are however semi-metals for which $\rho_{300K} \approx 0.1$ to 0.3 mΩ-cm, comparable to those reported for single-layer $BaFe_2As_2$ iron-arsenide superconducting parent ($\rho_{ab,300K}$=0.5 mΩ-cm) [33,34]. The plot of heat capacity in form of $C/T$ versus $T^2$ at low temperature is shown in the inset. The Sommerfeld-coefficient, $\gamma$, for all the compounds are estimated to the range between 6 to 10 mJ.K$^{-2}$.mol$^{-1}$, which is also comparable to $BaFe_2As_2$ ($\gamma \approx 6$ mJ.K$^{-2}$.mol$^{-1}$). The temperature dependence of magnetic susceptibility results, χ(T), are shown in the applied field of 1 Tesla, in **Fig. 2**, bottom. Low field measurements give no indication for a superconducting diamagnetic behavior. For all compounds, Curie tails are seen at lower temperatures. Pauli paramagnetic temperature-independent χ behavior is seen for $TlCu_2S_2$ and $TlCu_6S_4$. For the double-layer compounds, the susceptibility values are higher $\chi_{300K} \approx 2.5$ or $8\times10^{-4}$ cm$^3$/mol for $TlCu_4Se_3$ or $TlCu_4S_3$, respectively, close to that of $BaFe_2As_2$, $\chi_{300K} \approx 7\times10^{-4}$ cm$^3$/mol [33,34]. Although there is a clear large feature in susceptibility below ~133 K for $BaFe_2As_2$ for the spin-density-wave antiferromagnetic transition, there are also small features in χ at ~ 240 K or ~175, respectively, for n=2 compounds. In order to ensure non-magnetism, we checked these samples with neutron diffraction and looked for potential signatures of ordering of the Cu spins. Within the resolution of the measurements, no significant scattering intensity was observed at low temperatures that could be attributed to a static magnetic order. Hence, both $TlCu_4S_3$ and $TlCu_4Se_3$ are also considered non-magnets and refined with the tetragonal *P4/mmm* space group all throughout the temperature range and the lattice parameters are summarized in **Table 1**. There is one anomaly seen between the two samples, in that for the Se sample, the *c*-axis shrinks with increasing temperature.

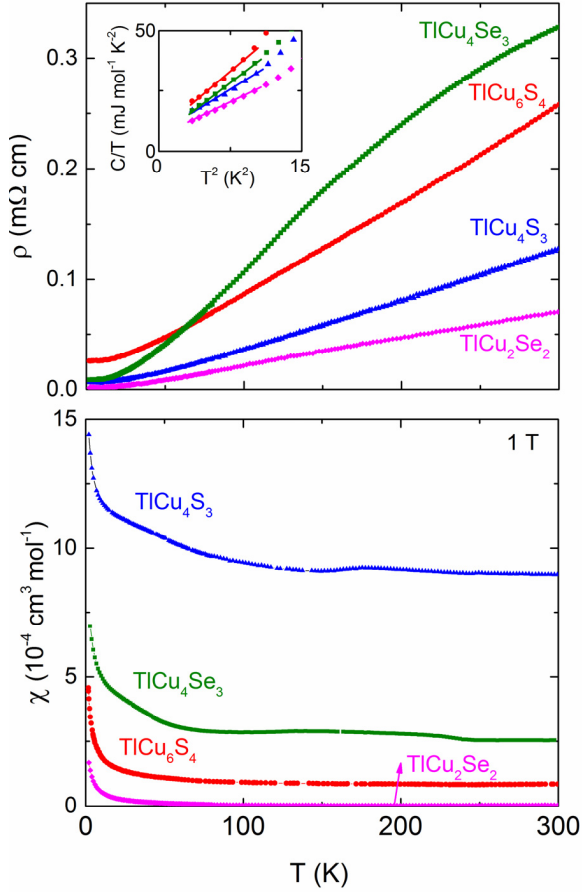

**Fig. 2.** Temperature-dependence of resistivity (top) and magnetic susceptibility (bottom) for $TlCu_2Se_2$, $TlCu_4S_3$, $TlCu_4Se_3$ and $TlCu_6S_4$. The inset shows $C/T$ versus $T^2$ along with linear fits at low temperatures.

**Table 1**: Lattice parameters for double-layered Tl-Cu-S structures. Data at 300 K are refined from X-ray diffraction data; lower temperature data are neutron diffraction results.

|      | $TlCu_4S_3$ |           | $TlCu_4Se_3$ |           |
|------|-------------|-----------|--------------|-----------|
| T(K) | $a$ (Å)     | $c$ (Å)   | $a$ (Å)      | $c$ (Å)   |
| 4    | 3.8642(2)   | 9.2943(4) | 3.9087(3)    | 9.9517(8) |
| 50   | 3.8653(2)   | 9.2948(5) | 3.9120(3)    | 9.9516(8) |
| 100  | -           | -         | 3.9225(3)    | 9.9406(8) |
| 250  | 3.8894(1)   | 9.3182(3) | 3.9641(4)    | 9.842(1)  |
| 300  | 3.894(1)    | 9.325(1)  | 3.976(1)     | 9.837(2)  |
|      | $TlCu_2Se_2$ |          | $TlCu_6S_4$  |           |
| 300  | 3.857(1)    | 14.041(1) | 3.935(1)     | 24.178(9) |

First principles density-functional theory calculations of the electronic structure of $TlT_2Se_2$ and $TlT_4Se_3$ ($T$ = Cu or Fe) are shown in **Fig. 3**. For both copper-based compounds, the majority of the copper DOS (indicated by the blue dot-dashed line) lies significantly below the Fermi level, which falls on a tail of the DOS curve. For the $TlCu_2Se_2$ structure (bottom), the Cu Fermi-level DOS $N_0$ is just 0.89 (eV-u.c.)$^{-1}$, or 0.45 per Cu atom (per eV), while the corresponding value for the $TlCu_4Se_3$ compound is 0.36 per Cu atom (per eV). These DOS values are well below what would be required to drive a magnetic instability, as given by the Stoner criterion $IN_0 > 0$, where $I$ is the exchange correlation integral, generally on the order of 0.4-0.5 eV for metallic Cu [35]. It is instructive to compare these results with those generated for $TlFe_2Se_2$ and the hypothetical $TlFe_4Se_3$ (insets of the figures). Here the *3d* atom DOS manifold has moved up much closer to the Fermi level, with regions of very high DOS very near $E_F$ (although $E_F$ itself falls within a narrow gap-like region for both prospective materials). These high DOS levels are consistent with a greater likelihood of magnetism for these materials, and bear some similarity to the parent compounds of the iron-based materials, which themselves exhibit magnetic instabilities.

Returning to the copper-based compounds, **Table 2** presents calculated squared plasma frequencies (in $eV^2$) for these materials. These squared frequencies are significantly greater for the $BaFe_2As_2$. This is consistent with the generally lower resistivity recalling that near or above the Debye temperature, the electron-phonon determined resistivity is inversely proportional to the squared plasma frequency. It is sensible that for both materials, the planar plasma frequencies are substantially larger than the *c*-axis values, as we expect the Tl spacer layer to reduce *c*-axis transport. This is particularly expected given the general lack of Tl character in the DOS in **Fig. 3**. It is of interest to note that these materials have essentially the same ratio of planar to *c*-axis conductivity. One might perhaps expect double-layer materials to be more three dimensional than single-layer, since more of the volume is comprised of the CuSe layer, which contains significant *c*-axis directional bonding and thus might be expected to encourage *c*-axis transport, but it appears here that the intervening Tl layer predominates this effect.

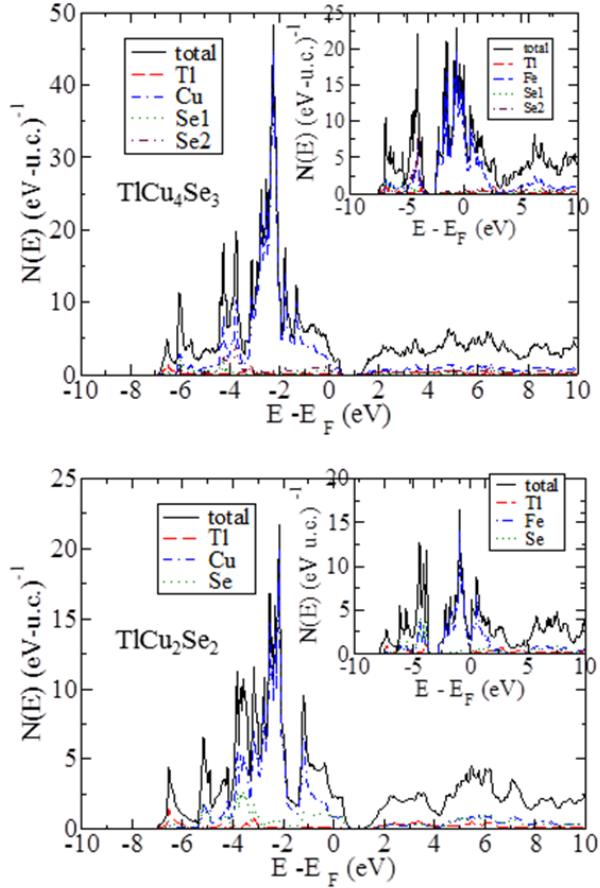

**Fig. 3:** The calculated densities-of-states of $TlCu_4Se_3$ (top) and $TlCu_2Se_2$ (bottom). Insets depict the calculated densities-of-states with the copper replaced by iron.

**Table 2.** The calculated plasma frequencies of the $TlCu_2Se_2$ and $TlCu_4Se_3$.

| Material | $\sigma_\parallel$ (eV$^2$) | $\sigma_c$ (eV$^2$) | $\sigma_\parallel/\sigma_c$ |
|---|---|---|---|
| $TlCu_2Se_2$ | 13.0 | 4.7 | 2.78 |
| $TlCu_4Se_3$ | 7.1 | 2.6 | 2.68 |

All the Tl (or alkali metal)-based cooper-chalcogenides of $ACu_{2n}Ch_{n+1}$ exhibit bad metallic behaviors and most of show weak susceptibility dependence in the whole temperature range [36-40]. The lack of superconducting/magnetic behavior is in contrast to the spectacular superconductivity in the cuprates, with $T_c$ values as high as 165 K under pressure. Although the cuprates are still not understood, one can reasonably argue that the essential presence of oxygen predisposes the parent compounds to a Mott insulating state, unlike the Cu-based compounds studied here. This means that in the cuprates upon doping at low carrier concentrations (compared to the metallic Cu compounds) there is still remaining magnetic character, with only partial screening, which ultimately drives the superconductivity. Selenium and even sulfur are not nearly as electronegative as oxygen. Thus the cuprates remain as a striking anomaly to the more typical non-superconducting behavior of copper-based compounds. For the metallic copper compounds measured here, there is simply a lack of Fermi-level spectral weight to drive a magnetic instability as in the iron-based materials, and semiconducting copper compounds do not have the magnetic character which upon doping could lead to a superconducting instability. It is also of interest that no one has succeeded in finding high-temperature superconductivity in nickelates analogues of the cuprates, despite substantial efforts in this direction. There is indeed something highly unusual about the particular combination of copper and oxygen that leads to high-temperature superconductivity in the cuprates.

**Acknowledgement**

This work was supported by the U. S. Department of Energy (DOE), Office of Science, Basic Energy Sciences, Materials Science and Engineering Division (L.L, D.P, A.S). The work at ORNL's High Flux Isotope Reactor (HFIR) was sponsored by the Scientific User Facilities Division, Office of BES, U.S. DOE (C.D.).